# Understanding the magnetic resonance spectrum of nitrogen vacancy centers in an ensemble of randomly-oriented nanodiamonds

Keunhong Jeong [*,1,2], Anna J. Parker [*,1,2], Ralph H. Page[1], Alexander Pines[1,2], Christophoros C. Vassiliou[1,2], Jonathan P. King[†,1,2]

1. Department of Chemistry, University of California, Berkeley.
2. Materials Sciences Division, Lawrence Berkeley National Laboratory

*These authors contributed equally to this work*

† jpking@berkeley.edu

*Supporting Information Placeholder*

Nanodiamonds containing nitrogen vacancy (NV⁻) centers show promise for a number of emerging applications including targeted *in vivo* imaging and generating nuclear spin hyperpolarization for enhanced NMR spectroscopy and imaging. Here, we develop a detailed understanding of the magnetic resonance behavior of NV⁻ centers in an ensemble of nanodiamonds with random crystal orientations. Two-dimensional optically detected magnetic resonance spectroscopy reveals the distribution of energy levels, spin populations, and transition probabilities that give rise to a complex spectrum. We identify overtone transitions that are inherently insensitive to crystal orientation and give well-defined transition frequencies that access the entire nanodiamond ensemble. These transitions may be harnessed for high-resolution imaging and generation of nuclear spin hyperpolarization. The data are well described by numerical simulations from the zero- to high-field regimes, including the intermediate regime of maximum complexity. We evaluate the prospects of nanodiamond ensembles specifically for nuclear spin hyperpolarization and show that frequency-swept dynamic nuclear polarization may transfer a large amount of the NV⁻ center's hyperpolarization to nuclear spins by sweeping over a small region of its spin spectrum.

The nitrogen vacancy (NV⁻) center in diamond, with its stable fluorescence, spin-1 ground state and optical spin polarization and readout, enables many emerging applications such as high spatial resolution sensing and magnetic resonance [1-9], solid-state qubits[10-12], and nuclear spin hyperpolarization[13-23]. In particular, biocompatible nanodiamonds are of interest for *in vivo* applications such as fluorescence-detected biomedical imaging[24,25] and as magnetic resonance contrast agents[26,27]. Owing to their high surface area, nanodiamonds have also been proposed as a general source for nuclear spin hyperpolarization transfer to target molecules for magnetic resonance signal enhancement[18,20]. All of these applications rely on the ability to access magnetic resonance transitions of the NV⁻ centers. However, since the NV⁻ spin properties depend strongly on their orientation with respect to an external magnetic field[28,29], the use of nanodiamond ensembles poses significant challenges.

Neglecting the weak hyperfine interaction with nearby $^{14}$N and the rare $^{13}$C nuclei, the electron spin Hamiltonian of NV⁻ center is:

$$H = D\left(S_z^2 - \frac{1}{3}S^2\right) + \gamma_e B_0 (S_x \sin\theta + S_z \cos\theta),$$

Where **D** = 2,870 MHz is zero field splitting, $\gamma_e$ = 2.8 MHz/G is the electron gyromagnetic ratio, and θ is the angle between the magnetic field and the NV⁻ axis (Figure 1a).

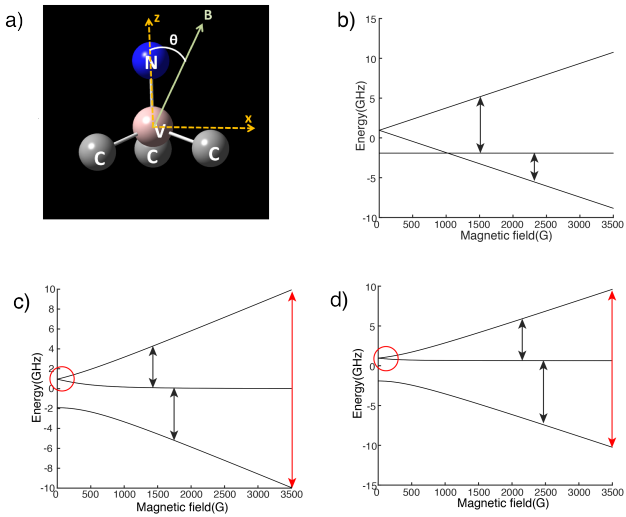

**Figure 1. NV⁻ center orientation and energy levels**. a) The NV⁻ center C$_{3v}$ symmetry axis forms an angle θ with respect to an externally applied magnetic field. Energy levels as a function of magnetic field of an NV⁻ center whose axis is oriented (b) 0°, (c) 54.74°, and (d) 109.5° with respect to the magnetic field. In (b), single quantum transitions are allowed (black arrows), but overtone transitions are forbidden. In (c) and (d), mixing of states by the transverse component of the Zeeman interaction allows overtone transitions (high-field overtones shown as red arrows and the region of low-field overtones are indicated by red circles).

Up to three transitions may be observed between the three spin states of the NV⁻ center (Fig. 1b-c). The relative intensity of the ODMR signal for a given transition is proportional to a factor $\kappa^{30}$, where:

$$\kappa = |\langle\varphi_f|\gamma_e(S_x + S_y)|\varphi_i\rangle|^2 (\Delta\langle\rho\rangle)(\Delta\langle S_z^2\rangle)$$
$$\Delta\langle\rho\rangle = \langle\varphi_f|\rho|\varphi_f\rangle - \langle\varphi_i|\rho|\varphi_i\rangle$$
$$\Delta\langle S_z^2\rangle = \langle\varphi_f|S_z^2|\varphi_f\rangle - \langle\varphi_i|S_z^2|\varphi_i\rangle.$$

$\langle\varphi_f|\gamma_e(S_x + S_y)|\varphi_i\rangle$ is the matrix element for a magnetic dipole transition between an initial and final eigenstate, assuming the microwave field is perpendicular to the z-axis (see supporting information). $\Delta\langle\rho\rangle$ is the population difference of the two eigenstates involved in the transition, and $\Delta\langle S_z^2\rangle$ is proportional to the difference in fluorescence intensity between the initial and final states. In our simulations, the density operator ρ was taken to be $\rho = E - S_z^2$, where E is the identity operator, which assumes 100% population of the m$_s$ = 0 state during optical pumping.

Continuous-wave ODMR spectra were acquired by detecting fluorescence while applying an amplitude-modulated microwave field. The samples were mounted in an electromagnet and the ODMR spectra were acquired at magnetic fields ranging from 0 to 350 mT to create two-dimensional spectra. The microwave field was created at the sample with a 3 mm wire loop and optical excitation at 532 nm was provided at a power of 1 Watt by a Coherent Verdi G15 laser with a beam waist of 1.5 mm. The signal corresponds to the in-phase component of the photodiode voltage detected at the modulation frequency (see supporting info. for more details).

The single crystal sample is a 2 x 2 x 0.2 mm [1 0 0] surface oriented synthetic diamond produced via a high-pressure high-temperature (HPHT) synthesis (Sumitomo Electric Carbide Inc.). NV⁻ centers were formed at a density of 7.8 ppm by irradiation with 1 MeV electrons and annealing at 800 °C for 2 hours. When a [1 1 1] axis is aligned with the magnetic field, a single crystal exhibits two sets of transitions: a set of two transitions resulting from NV⁻ centers with symmetry axes aligned with the field and a set of three from those misaligned by 109.5°. (Fig. 2a). When a [1 0 0] axis is aligned with the field, all NV⁻ axes form an angle of 54.7° with the field and a single set of three transitions is observed (Fig. 2b). Transitions associated with NV⁻ centers misaligned with the magnetic field exhibit a nonlinear dependence on magnetic field, most apparent below 1000 G. One of the transitions of each misaligned defect has an apparent gyromagnetic ratio twice that of the NV⁻ center and diminishes in intensity at higher fields. These are $\Delta m = \pm 2$ "overtone" transitions that, while nominally forbidden, become allowed by the mixing of states induced by the zero-field splitting term in the Hamiltonian[31,32]. In the low-field regime a second type of overtone is barely observed at magnetic fields below 1000 G, where the quantization axis is now defined by the defect symmetry axis and the NV⁻ Zeeman interaction is the perturbation. This overtone tends to zero frequency at zero-field. In all cases, the data were well fit by simulations (Fig. 2c and 2d) based on the physical model presented above (details of numerical simulation are presented in the supporting information). The energy levels involved in the overtone transitions are labeled in Figure 1c and 1d.

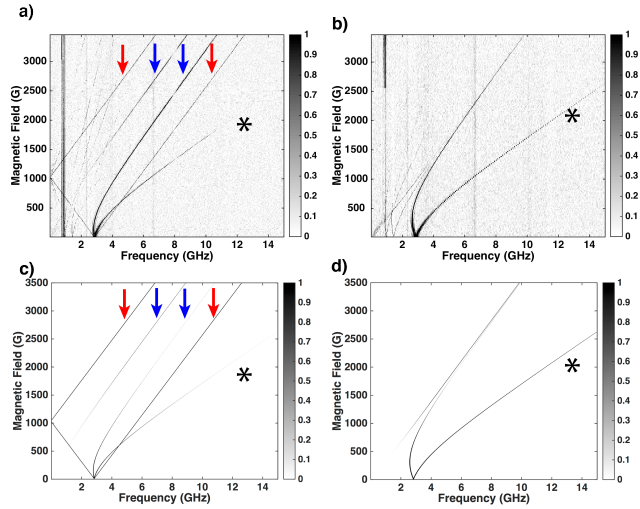

**Figure 2. Orientation-dependent spin transitions of NV⁻ centers.** 2D ODMR with the magnetic field aligned along the a) [1 1 1] and b) [1 0 0] axes. Red arrows indicate spectra from 0° oriented NV⁻ centers which exhibit an avoided crossing around 1000 G. Blue arrows indicate a 109.5° orientation. Overtone transitions are indicated by an asterisk. The vertical bands are instrument artifacts (see Methods). Lines in the data not reproduced by the simulations result from harmonics of the microwave amplifier. c) and d) show results for numerical simulations corresponding to the results in a) and b), respectively. For figures a-d, the intensity scales represent ODMR contrast are normalized to 0.3, 0.001, 0.001, and 0.001, respectively.

2D ODMR of nanodiamond powder was measured using the same method as for single crystals (Figure 3a,b). The powder sample was 10 mg of 100 nm fluorescent diamond nanoparticles (Adamas Nanotechnologies Inc.) For intermediate- and high-fields, the ODMR spectrum exhibited several GHz wide powder pattern, with corresponding reduction in ODMR contrast. At high fields, the overtone transitions converge in frequency, creating a sharp peak.

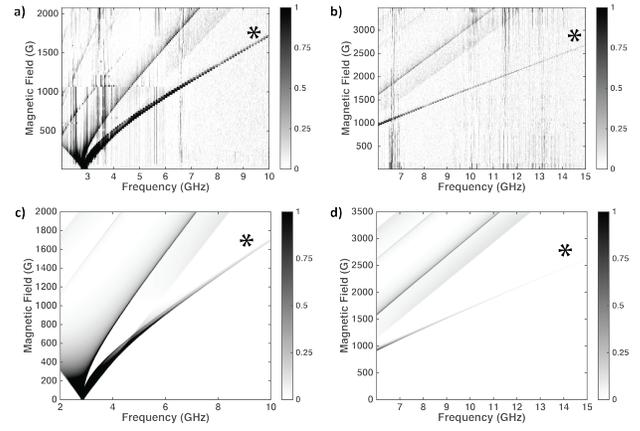

**Figure 3. 2D ODMR of randomly oriented nanodiamond powders**. a), b) 2D ODMR of NV⁻ centers in a randomly oriented powder of 100nm diamond crystallites. Overtone transitions are indicated with an asterisk. The vertical bands in (a) and (b) are instrument artifacts (see Methods). c), d) 2D ODMR simulation of NV⁻ centers in diamond powder corresponding to the data in (a) and (b). Simulations are normalized to match the intensity scale of the experimental data.

To further investigate the high-field overtone transitions, we obtained high signal-to-noise one-dimensional (1D) ODMR spectra at specific magnetic fields (985 G and 1313 G) (Figure 4). These transitions exhibit characteristic powder pattern shapes[33,34] that are much narrower than the main region of the spectrum and narrow as the field is increased. Additionally, these spectra show some narrow features and structure not captured by the model. Such features could arise from an anisotropic distribution of crystal orientations, orientation-dependent scattering or reflection by individual crystallites, or orientation-dependent photodynamics that affect the populations of the energy levels and ODMR contrast[35].

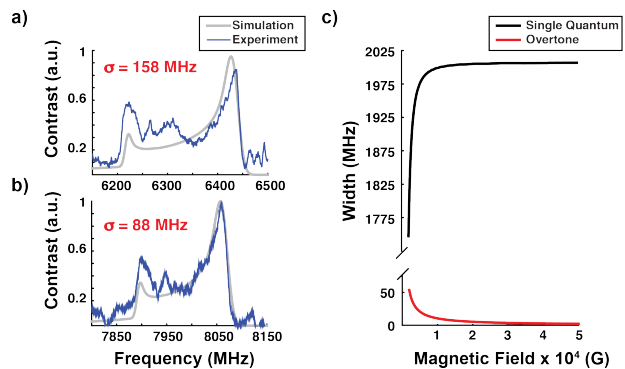

**Figure 4. Narrowing of overtone transitions in high-field limit.** 1D spectra of overtone transitions at a) 985 G and b) 1313 G. The nonzero widths of the overtone spectra are attributed to orientation dependent $2^{nd}$ and higher-order contributions from the zero-field splitting Hamiltonian. The main features of the spectra are well fit by simulations (gray lines). Spectra in a) and b) were fit using the procedure described in the Methods Section and are normalized to 4.0 x $10^{-5}$ and 1.6 x $10^{-5}$, respectively. The simulated characteristic width of the single quantum and overtone transitions is given in (c) as a function of magnetic field. In the high field limit, σ approaches approximately 2 GHz for the single quantum powder pattern and 0 for the overtone powder pattern. Details on the calculation of the characteristic width are given in the supporting information.

Magnetic dipole transitions have selection rules of $\Delta m = \pm 1$ or $\Delta m = 0$ for transverse and longitudinal irradiation, respectively. However, when the crystal axis and magnetic field are not aligned, $m$ is no longer a good quantum number and all transitions are allowed. We can roughly consider the overtone transition in three regions. Near zero field there exists a $\Delta m \cong \pm 2$ transition as defined by the NV⁻ symmetry axis. This transition tends towards zero frequency and zero intensity as the magnetic field goes to zero. At intermediate fields, $D \cong \gamma B$, the energy levels are highly mixed and no distinct overtone transition is observed. At high field the quantization axis becomes the external field and, again, a $\Delta m \cong \pm 2$ transition is observed this time with effective gyromagnetic ratio of $2\gamma$. In the high-field regime, the overtone transition frequency is independent of crystal orientation to a first-order approximation, and the powder pattern observed for these transitions is much narrower than that for the $\Delta m \cong \pm 1$ transitions. The nonzero width of the overtone spectra observed in Fig. 4 can be attributed to second- and higher-order corrections to the energy levels and narrows as these terms are suppressed at higher fields as also seen in the simulations. The magnetic field dependence of the characteristic width, σ, of the two distinct powder patterns is illustrated fully in Figure 4c. We define σ as the square root of the second moment of the powder pattern. In the high-field limit at 5 x $10^4$ G, σ approaches approximately 2 GHz for the single-quantum powder pattern and 0 for the overtone powder pattern. Details on the calculation of the characteristic width are given in the supporting information.

Two proposed applications of hyperpolarization with NV⁻ centers are: the enhancement of NMR signal from target molecules, and the use of the diamonds themselves as MRI contrast agents. The former will require high surface area afforded by diamond nanocrystals and the latter requires nanocrystals for introduction into the body. These applications differ in that the target nuclei are either inside or outside of the diamond, but both cases demand a strategy for addressing the orientation-dependent spin physics of NV⁻ centers. Here we use the insight gained from the present simulations and experiments to evaluate the prospects for hyperpolarization of randomly-oriented diamond powders. An established method for polarizing nuclei from an inhomogenously-broadened electron spin spectrum is the integrated solid effect (ISE[36,37]), which involves quasi-adiabatic field or frequency sweeps to coherently transfer polarization from electron to nuclear spins. ISE has been suggested as a method for obtaining hyperpolarization in nanodiamond powder, but difficulties in obtaining GHz-wide frequency sweeps have thus far prevented its implementation[18]

Since the NV⁻ spin level populations are orientation dependent, it is not immediately clear that ISE, which effectively integrates the NV⁻ polarization for some region of the spectrum, can generate nuclear spin hyperpolarization in randomly-oriented powders. The available NV⁻ polarization for a given transition is the population difference $\Delta\langle\rho\rangle$. Using the model developed here, the available NV⁻ center polarization density as a function of field and frequency is shown in Figure 5 (a and b). The polarization density represents the number of nuclear spins that may be polarized per unit frequency in a sweep. This is of practical importance in the design of hyperpolarization experiments since the range of a frequency sweep is often limited by the bandwidth of a microwave cavity[18] and fast magnetic field sweeps over large ranges are limited by sweep coil inductance. We can see that the polarization density is concentrated in frequency regions corresponding to NV⁻ centers at a 90° orientation and also to the high-field overtone transitions. In Figures 5c and 5d we plot the maximum polarization that can be generated as a function of limited sweep width for both the single-quantum and overtone regions of the spectrum. We see that, at a field of 2000 G, a sweep width of ~3 GHz gives the maximum integrated polarization of about 50% from the single quantum transitions. However, a sweep width of only 100 MHz gives a polarization of ~10%. The overtone transitions have even more concentrated polarization density where a sweep of ~200 MHz gives ~25% polarization. These results are encouraging for the prospect of hyperpolarization since the indicate that hyperpolarization may be generated with sweep widths on the order of 100 MHz, rather than several GHz, bringing the sweep width within the bandwidth of an X-band electron paramagnetic resonance cavity with a quality factor of 100.

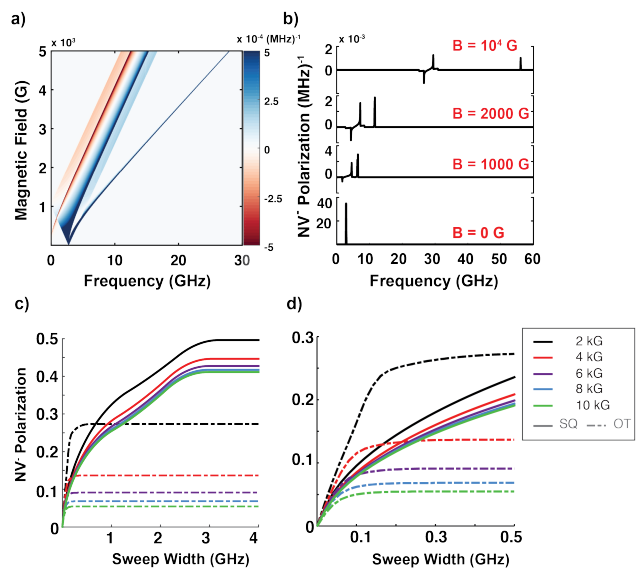

**Figure 5 Prospects for hyperpolarization from randomly oriented powders.** (a) 2D plot of the NV⁻ polarization density as a function of field and frequency. 1D slices in (b) show concentrated regions of polarization density at various magnetic fields. The units of the amplitude in (a) and (b) are dimensionless polarization density per unit frequency, $(MHz)^{-1}$. (c) Maximum integrated NV⁻ polarization as a function of frequency sweep width for both the overtone (OT) and single quantum (SQ) powder patterns at several magnetic fields. (d) Expanded view of integrated NV⁻ polarization at sweep widths below 0.5 GHz. The maximum integrated NV⁻ polarization converges in the high field limit for both powder patterns, with the overtone powder pattern converging more slowly than the SQ powder pattern. At magnetic fields at 8 kG and above, sweeping over the SQ powder pattern is always advantageous.

We also point out that in fluorescence imaging of nanodiamonds, magnetic resonance control of NV⁻ spins enhances contrast and can provide an additional degree of spatial information through the application of magnetic field gradients[38]. Determining the imaging point spread function in this case is equivalent to calculating the ODMR spectrum[39]. Understanding the ensemble ODMR spectrum is critical to extending ODMR imaging to stronger gradients and higher fields where narrow overtone transitions will provide enhanced resolution.

In summary, 2D ODMR spectra of NV⁻ centers in diamond ranging from the zero- to high-field regimes in single crystals and powders are well described by a model accounting for the angular dependence of transition frequencies, spin polarization, and optical contrast. While the NV⁻ spectrum can span almost 6 GHz, the polarization density is concentrated in a few regions corresponding to the 90° orientation and the overtone transitions. This concentration of polarization density affords the possibility of frequency-swept DNP techniques to achieve nuclear spin hyperpolarization in high surface area nanodiamonds. Finally, the detailed knowledge provided by the model enables the encoding of spatial information in targeted imaging applications involving nanodiamonds.